\documentclass[12pt]{article}

\usepackage{hyperref}
\usepackage{cite}
\usepackage{units}
\usepackage{epsfig}
\usepackage{pifont}
\usepackage{amsmath,amsfonts,amssymb,amsthm,mathrsfs}
\usepackage{graphicx,chngcntr,slashed}
\usepackage{cancel}
\usepackage{pstricks}
\usepackage{afterpage}
\usepackage{braket}
\usepackage{caption}
\usepackage{subcaption}
\usepackage{xcolor}
\usepackage{makecell}
\usepackage{physics}
\usepackage{multirow}

\newcommand{\dslash}[1]{\slash\!\!\! #1}

\def\mathswitch#1{\relax\ifmmode#1\else$#1$\fi}

\newcommand{\zp}{Z^{\prime}}
\newcommand{\be}{\begin{equation}}
\newcommand{\ee}{\end{equation}}
\newcommand{\bea}{\begin{eqnarray}}
\newcommand{\eea}{\end{eqnarray}}

\newcommand{\mycaption}[1]{\caption{\sl #1}}

\makeatletter
\def\section{\@startsection {section}{1}{\z@}{+3.0ex plus +1ex minus
  +.2ex}{2.3ex plus .2ex}{\large\bf\boldmath}}
\def\subsection{\@startsection{subsection}{2}{\z@}{+2.5ex plus +1ex
minus +.2ex}{1.5ex plus .2ex}{\normalsize\bf\boldmath}}
\def\subsubsection{\@startsection{subsubsection}{3}{\z@}{+3.25ex plus
 +1ex minus +.2ex}{1.5ex plus .2ex}{\normalsize\it}}

\graphicspath{{.}{plots/}}

\begin{document}
\thispagestyle{empty}

\def\thefootnote{\fnsymbol{footnote}}

\begin{flushright}
\end{flushright}

\vspace{1cm}

\begin{center}

{\Large {\bf Exploring the SMEFT at dimension-8 with Drell-Yan transverse momentum measurements}}
\\[3.5em]
{\large
Radja~Boughezal$^1$, Yingsheng Huang$^{1,2}$ and Frank~Petriello$^{1,2}$ 
}

\vspace*{1cm}

{\sl
$^1$ HEP Division, Argonne National Laboratory, Argonne, Illinois 60439, USA \\[1ex]
$^2$ Department of Physics \& Astronomy, Northwestern University,\\ Evanston, Illinois 60208, USA
}

\end{center}


\begin{abstract}
  
  We demonstrate that measurements of the neutral-current Drell-Yan transverse momentum distribution binned in invariant mass are sensitive to unexplored dimension-8 parameters of the Standard Model Effective Field Theory (SMEFT). These distributions are sensitive to four-fermion operators with additional QCD field strength tensors. The determination of the Wilson coefficients of these operators provides a useful diagnostic tool that distinguishes possible ultraviolet completions of the SMEFT. We study how well these effects can be probed by current LHC data, and explore the sensitivity of the future high-luminosity LHC (HL-LHC) to these operators. We find that the HL-LHC data has the potential to strongly probe this sector of the SMEFT.
      
\end{abstract}

\setcounter{page}{0}
\setcounter{footnote}{0}

\newpage


\section{Introduction}

The Standard Model (SM) of particle physics successfully describes phenomena ranging from low-energy nuclear physics to high-energy  collisions. However, since it does not contain neutrino masses nor dark matter, and cannot explain certain observations such as the matter-antimatter asymmetry in the universe, undiscovered physics beyond the SM that explains these mysteries must exist. Experiments at the Large Hadron Collider (LHC) and elsewhere are probing the SM at the TeV scale, searching for solutions to these outstanding problems. Since no conclusive deviation from SM predictions has yet been found, a major theme of current research is to understand how heavy new physics can be indirectly probed and constrained by available and upcoming data. This effort helps guide searches for new physics by suggesting in what channels measurable deviations from SM predictions may occur given the current bounds. In the event of a discovery it would also indicate what measurements can serve as diagnostic tools to distinguish between different models of new physics.

A convenient theoretical framework for investigating indirect signatures of heavy new physics is the SM Effective Field Theory (SMEFT). The SMEFT is formed by adding higher-dimensional operators to the SM Lagrangian that are consistent with the SM gauge symmetries and formed only from SM fields. The higher-dimensional operators in the SMEFT are suppressed by appropriate powers of a characteristic energy scale $\Lambda$ below which heavy new fields are integrated out. The SMEFT encapsulates a broad swath of new physics models, making it easier to simultaneously study numerous theories without focusing on details of their ultraviolet completions that do not matter at low energies. The use of the SMEFT framework to analyze LHC data is similar in spirit to the use of the $S$ and $T$ parameters at LEP to bound entire classes of new physics models, and the global fitting of SMEFT parameters at the LHC promises to provide as powerful probe of beyond the SM theories as the global electroweak precision fit did at LEP. Complete, non-redundant bases for the dimension-6~\cite{Buchmuller:1985jz,Arzt:1994gp,Grzadkowski:2010es} and dimension-8 operators~\cite{Murphy:2020rsh,Li:2020gnx} have been constructed. Odd-dimensional operators violate lepton-number and are not considered here.
It is an ongoing effort to analyze the numerous available data within the SMEFT framework, primarily in partial analyses of individual SMEFT sectors~\cite{Han:2004az,Cirigliano:2012ab,Chen:2013kfa,Ellis:2014dva,Wells:2014pga,Falkowski:2014tna,Cirigliano:2016nyn,deBlas:2016ojx,Hartmann:2016pil,Falkowski:2017pss,Alioli:2017ces,Alioli:2017nzr,Alioli:2018ljm,Biekotter:2018rhp,Grojean:2018dqj,Baglio:2020oqu,Boughezal:2020uwq,Boughezal:2020klp,Ricci:2020xre,Horne:2020pot,Boughezal:2021kla,Ethier:2021ydt,Boughezal:2022pmb}. Recent work has been devoted to performing a global, simultaneous fit of all data available~\cite{Pomarol:2013zra,DiVita:2017eyz,Almeida:2018cld,Ellis:2018gqa,Hartland:2019bjb,Brivio:2019ius,vanBeek:2019evb,Aoude:2020dwv,Ellis:2020unq,Dawson:2020oco,Greljo:2021kvv,Ethier:2021bye},
and to study the interplay between SMEFT fits and the extraction of parton distributions from data \cite{Carrazza:2019sec,Greljo:2021kvv}.

Our focus in this manuscript is on semi-leptonic four-fermion operators in the SMEFT. These coefficients are not constrained by current global fits of top quark, Higgs boson and electroweak data~\cite{Ellis:2020unq,Ethier:2021bye}. While they can be probed by low-energy data~\cite{Falkowski:2017pss}, the strongest bounds come from Drell-Yan data at the LHC. Previous results have shown that existing Drell-Yan data is precise enough to probe dimension-8 operators in the SMEFT~\cite{Alioli:2020kez,Boughezal:2021tih,Kim:2022amu}. These works focused on measurements of the invariant mass distribution of the lepton pair in the Drell-Yan process. A  motivation of our paper is to demonstrate that LHC data sets not originally intended as new physics searches can be sensitive to unprobed regions of the SMEFT parameter space, and therefore have unexpected sensitivity to physics beyond the SM. In particular we focus on the recent CMS measurement of the doubly-differential distribution of invariant mass and transverse momentum in the Drell-Yan process~\cite{CMS:2022ubq}, intended as a probe of QCD dynamics. The measurement of transverse momentum makes this data set sensitive to partonic processes containing emission of gluons. These gluons can either be radiated from external legs, or directly from some heavy state that carries QCD color. In this second case they match to semi-leptonic four-fermion operators with an additional QCD field-strength tensor. Such operators first appear at dimension-8 in the SMEFT and are unconstrained by other data sets.

To illustrate our results we focus on a representative example in which only operators containing right-handed fields have non-zero Wilson coefficients. In this scenario our parameter space consists of three categories of operators: a dimension-6 four-fermion operator, two momentum-dependent dimension-8 operators that grow with energy and that have been considered in previous work~\cite{Alioli:2020kez,Boughezal:2021tih}, and a single CP-even semi-leptonic four-fermion operator with a gluon field-strength tensor that we henceforth label a gluonic operator. We show that a joint measurement of invariant mass and transverse momentum allows the gluonic operator to be probed independently of the other operators, as it has a distinct dependence on transverse momentum. We stress that the determination of Wilson coefficients for all three operator categories provides a useful diagnostic tool that distinguishes possible ultraviolet (UV) completions of the SMEFT. Although our primary interest is in the bottom-up analysis of the possible SMEFT parameter space, we consider the matching of example $Z^{\prime}$ and vector leptoquark states to this sector of the SMEFT, and show that they lead to very different patterns of Wilson coefficients for these three operator categories. Although current LHC data provides only weak constraints on the gluonic operator, we study the potential of the high-luminosity LHC (HL-LHC) to probe these effects, and find that significant bounds on all three categories of effects can be obtained. We encourage this measurement to be performed again as larger LHC data sets become available.

Our paper is organized as follows. We review in Section~\ref{sec:smeft} details of the SMEFT needed for our analysis. In Section~\ref{sec:UV} we study the matching of example UV states onto the four-fermion sector of the SMEFT. Our emphasis in this section is to show that very different patterns of Wilson coefficients can be obtained from different UV states, motivating the measurement of all possible operator types. In Section~\ref{sec:doubdiff} we show that the doubly-differential distribution in invariant mass and transverse momentum can simultaneously probe both the regular and gluonic semi-leptonic four-fermion operators. We perform fits to the current data in Section~\ref{sec:current}, and to simulated HL-LHC data in Section~\ref{sec:hllhc}. We conclude in Section~\ref{sec:conc}.

\section{Review of the SMEFT} \label{sec:smeft}

We review in this section aspects of the SMEFT relevant for
our analysis of the Drell-Yan process.  The SMEFT is an effective field theory extension of the SM that includes terms
suppressed by an energy scale $\Lambda$. Beyond this scale the ultraviolet completion of the EFT 
becomes important, and new particles beyond the SM appear.  In our study we keep terms through dimension-8 in the $1/\Lambda$ expansion, and 
ignore operators of odd-dimension which violate lepton number. Our Lagrangian becomes
\begin{equation}
{\cal L} = {\cal L}_{SM}+ \frac{1}{\Lambda^2} \sum_i C^{(6)}_{i} {\cal
  O}^{(6)}_{i} + \frac{1}{\Lambda^4} \sum_i C^{(8)}_{i} {\cal
  O}^{(8)}_{i} + \ldots,
\end{equation}
where the ellipsis denotes operators of higher dimensions.  The Wilson
coefficients defined above are dimensionless. Cross sections computed through ${\cal O}(1/\Lambda^4)$ will have contributions 
from the square of dimension-6 operators, as well as interferences between dimension-8 operators and the SM.

The categories of operators contributing to the Drell-Yan process through dimension-8 were extensively cataloged in Ref.~\cite{Boughezal:2021tih}. At the dimension-6 level three categories of operators contribute: corrections to the three-point 
vertices of gauge bosons with fermions, four-fermion operators, and dipole operators coupling fermions to gauge bosons. The vertex 
corrections lead to effects that scale with energy as ${\cal O}(v^2/\Lambda^2)$, where $v$ denotes the Higgs vacuum expectation value. 
These are subleading at high energies compared to the four-fermion operators that scale as ${\cal O}(s/\Lambda^2)$, and are strongly 
constrained by $Z$-pole observables~\cite{Dawson:2018dxp}. We therefore neglect these terms in our analysis. We additionally assume minimal flavor violation for the structure of our Wilson coefficients. This assumption makes all dipole operators, as well as all scalar and tensor four-fermion operators, proportional to SM Yukawa couplings. These couplings are small for the processes considered here, and can be safely neglected. This leaves us with only vector-like four-fermion operators contributing at dimension-6. The contributing terms are summarized 
below in Table~\ref{tab:ffops}.
\begin{table}[h!]
\centering
\begin{tabular}{|c|c||c|c|}
\hline
${\cal O}_{lq}^{(1)}$ & $(\bar{l}\gamma^{\mu} l) (\bar{q}\gamma_{\mu} q)$ &  ${\cal O}_{lu}$ & $(\bar{l}\gamma^{\mu} l) (\bar{u}\gamma_{\mu}  u)$ 
  \\
  ${\cal O}_{lq}^{(3)}$ & $(\bar{l}\gamma^{\mu} \tau^I l)
                     (\bar{q}\gamma_{\mu} \tau^I l q)$ & ${\cal O}_{ld}$ & $(\bar{l}\gamma^{\mu} l) (\bar{d}\gamma_{\mu}  d)$   
  \\
  ${\cal O}_{eu}$ & $(\bar{e}\gamma^{\mu} e) (\bar{u}\gamma_{\mu}  u)$
                                                                     & ${\cal O}_{qe}$ & $(\bar{q}\gamma^{\mu} q) (\bar{e}\gamma_{\mu}  e)$             
  \\
  ${\cal O}_{ed}$ & $(\bar{e}\gamma^{\mu} e) (\bar{d}\gamma_{\mu}  d)$
                                                                     & &
  \\
  \hline
\end{tabular}
\mycaption{Dimension-6 four-fermion operators contributing to Drell-Yan  at leading order in the coupling constants.\label{tab:ffops}}
\end{table}
$q$ and $l$ denote left-handed quark and lepton doublets, while
$u$, $d$ and $e$ denote right-handed singlets for the up quarks, down
quarks and leptons, respectively.  $\tau^I$ denote the SU(2)
Pauli matrices. 

Several classes of operators contribute at the dimension-8 level. Considering first the leading-order four-fermion process, we again have 
corrections to the $f\bar{f}V$ vertices, four-fermion operators with Higgs insertions, and four-fermion operators with derivative insertions. The first two categories of operators were shown in Ref.~\cite{Boughezal:2021tih} to be negligible for reasonable values of the Wilson coefficients, consistent with their energy scaling: ${\cal O}(v^4/\Lambda^4)$ for the first category and ${\cal O}(s v^2/\Lambda^4)$ for the second. 
The four-fermion operators with derivative insertions scale as ${\cal O}(s/\Lambda^4)$ and are non-negligible. They are shown in Table~\ref{tab:ffd8}. We note that the type-II operators lead to novel angular dependence~\cite{Alioli:2020kez}, but vanish upon integration over angles up to small corrections due to acceptance cuts~\cite{Boughezal:2021tih}. While we discuss them when matching of specific UV examples to the SMEFT, we do not consider them in our numerical analysis since the distributions considered here show little sensitivity to these effects. A proposal for a series of angular measurements to probe these terms was given in~\cite{Alioli:2020kez,Li:2022rag}.
\begin{table}[h!]
\begin{center}
  \begin{tabular}{| c | c ||  c | c |}
  \hline
  \multicolumn{2}{|c||}{Type-I} & \multicolumn{2}{c|}{Type-II}\\
    \hline
     $\mathcal{O}^{(1)}_{l^2q^2D^2}$ & $D^\nu \left(\overline{l} \gamma^\mu l \right) D_\nu \left(\overline{q} \gamma_\mu q \right)$ & $\mathcal{O}^{(2)}_{l^2q^2D^2}$ & $\left(\overline{l} \gamma^{(\mu} \overset{\leftrightarrow}{D^{\nu)}} l \right) \left(\overline{q} \gamma_{\mu(} \overset{\leftrightarrow}{D_{\nu)}}q \right)$  \\

$\mathcal{O}^{(3)}_{l^2q^2D^2}$  &$D^\nu \left(\overline{l} \gamma^\mu\tau^I l \right) D_\nu \left(\overline{q} \gamma_\mu\tau^I q \right)$
    &$\mathcal{O}^{(4)}_{l^2q^2D^2}$ & $\left(\overline{l} \tau^I \gamma^{(\mu} \overset{\leftrightarrow}{D^{\nu)}} l \right) \left(\overline{q} \tau^I\gamma_{\mu(} \overset{\leftrightarrow}{D_{\nu)}}q \right)$ \\ 

$\mathcal{O}^{(1)}_{e^2u^2D^2}$  & $D^\nu \left(\overline{e} \gamma^\mu e \right) D_\nu \left(\overline{u} \gamma_\mu u \right)$
   & $\mathcal{O}^{(2)}_{e^2u^2D^2}$ & $\left(\overline{e} \gamma^{(\mu} \overset{\leftrightarrow}{D^{\nu)}} e \right) \left(\overline{u} \gamma_{\mu(} \overset{\leftrightarrow}{D_{\nu)}}u \right)$  \\ 

$\mathcal{O}^{(1)}_{e^2d^2D^2}$  & $D^\nu \left(\overline{e} \gamma^\mu e \right) D_\nu \left(\overline{d} \gamma_\mu d \right)$
    &$\mathcal{O}^{(2)}_{e^2d^2D^2}$ & $\left(\overline{e} \gamma^{(\mu} \overset{\leftrightarrow}{D^{\nu)}} e \right) \left(\overline{d} \gamma_{\mu(} \overset{\leftrightarrow}{D_{\nu)}}d \right)$ \\

$\mathcal{O}^{(1)}_{l^2u^2D^2}$ & $D^\nu \left(\overline{l} \gamma^\mu l \right) D_\nu \left(\overline{u} \gamma_\mu u \right)$
   &  $\mathcal{O}^{(2)}_{l^2u^2D^2}$ & $\left(\overline{l} \gamma^{(\mu} \overset{\leftrightarrow}{D^{\nu)}} l \right) \left(\overline{u} \gamma_{\mu(} \overset{\leftrightarrow}{D_{\nu)}}u \right)$\\ 

$\mathcal{O}^{(1)}_{l^2d^2D^2}$ &  $D^\nu \left(\overline{l} \gamma^\mu l \right) D_\nu \left(\overline{d} \gamma_\mu d \right)$
     & $\mathcal{O}^{(2)}_{l^2d^2D^2}$ & $\left(\overline{l} \gamma^{(\mu} \overset{\leftrightarrow}{D^{\nu)}} l \right) \left(\overline{d} \gamma_{\mu(} \overset{\leftrightarrow}{D_{\nu)}}d \right)$ \\

$\mathcal{O}^{(1)}_{q^2e^2D^2}$ & $D^\nu \left(\overline{q} \gamma^\mu q \right) D_\nu \left(\overline{e} \gamma_\mu e \right)$
   & $\mathcal{O}^{(2)}_{q^2e^2D^2}$ & $\left(\overline{e} \gamma^{(\mu} \overset{\leftrightarrow}{D^{\nu)}} e \right) \left(\overline{q} \gamma_{\mu(} \overset{\leftrightarrow}{D_{\nu)}}q \right)$\\
   \hline
  \end{tabular}
    \mycaption{Dimension-8 two-derivative four-fermion operators that contribute to the Drell-Yan process. The type-I operators affect inclusive invariant mass and transverse momentum distributions, while the type-II operators affect only angular distributions and vanish when integrated inclusively over angles. The parentheses in the superscripts denote symmetrization over the enclosed indices, while $\overset{\leftrightarrow}{D_{\nu}} = \overset{\rightarrow}{D_{\nu}}-\overset{\leftarrow}{D_{\nu}}$. \label{tab:ffd8}}
\end{center}
\end{table}

In our study we will consider the transverse momentum spectrum in Drell-Yan. In this case we also need to consider dimension-8 
operators with a gluon field-strength inserted. The possible operators of this form were enumerated in Refs.~\cite{Li:2020gnx,Murphy:2020rsh}. 
We organize them according to their $CP$ transformation properties in Table~\ref{tab:ffd8pT}. These transformation rules can be obtained using results from any standard QFT text, or from studying the structure of an explicit amplitude calculation. Since the $CP$-odd operators do not interfere with the tree-level SM amplitudes we do not consider them in this study.

\begin{table}[h!]
\begin{center}
  \begin{tabular}{| c | c ||  c | c |}
  \hline
  \multicolumn{2}{|c||}{$CP$-even} & \multicolumn{2}{c|}{$CP$-odd}\\
    \hline
	$\mathcal{O}^{(1)}_{l^2 q^2 \tilde{G}}$ & $(\bar{l}\gamma^{\mu} l) (\bar{q}\gamma^{\nu} T^A q) \tilde{G}^A_{\mu\nu}$ &
	$\mathcal{O}^{(1)}_{l^2 q^2 G}$ & $(\bar{l}\gamma^{\mu} l) (\bar{q}\gamma^{\nu} T^A q) G^A_{\mu\nu}$ \\	
	
	$\mathcal{O}^{(2)}_{l^2 q^2 \tilde{G}}$ & $(\bar{l}\tau^I \gamma^{\mu} l) (\bar{q}\tau^I\gamma^{\nu} T^A q) \tilde{G}^A_{\mu\nu}$ &
	$\mathcal{O}^{(2)}_{l^2 q^2 G}$ & $(\bar{l}\tau^I\gamma^{\mu} l) (\bar{q}\tau^I\gamma^{\nu} T^A q) G^A_{\mu\nu}$ \\	
	
	$\mathcal{O}_{e^2 u^2 \tilde{G}}$ & $(\bar{e}\gamma^{\mu} e) (\bar{u}\gamma^{\nu} T^A u) \tilde{G}^A_{\mu\nu}$ &
	$\mathcal{O}_{e^2 u^2 G}$ & $(\bar{e}\gamma^{\mu} e) (\bar{u}\gamma^{\nu} T^A u) G^A_{\mu\nu}$ \\
	
	$\mathcal{O}_{e^2 d^2 \tilde{G}}$ & $(\bar{e}\gamma^{\mu} e) (\bar{d}\gamma^{\nu} T^A d) \tilde{G}^A_{\mu\nu}$ &
	$\mathcal{O}_{e^2 d^2 G}$ & $(\bar{e}\gamma^{\mu} e) (\bar{d}\gamma^{\nu} T^A d) G^A_{\mu\nu}$ \\	
	
	$\mathcal{O}_{l^2 u^2 \tilde{G}}$ & $(\bar{l}\gamma^{\mu} l) (\bar{u}\gamma^{\nu} T^A u) \tilde{G}^A_{\mu\nu}$ &
	$\mathcal{O}_{l^2 u^2 G}$ & $(\bar{l}\gamma^{\mu} l) (\bar{u}\gamma^{\nu} T^A u) G^A_{\mu\nu}$ \\
	
	$\mathcal{O}_{l^2 d^2 \tilde{G}}$ & $(\bar{l}\gamma^{\mu} l) (\bar{d}\gamma^{\nu} T^A d) \tilde{G}^A_{\mu\nu}$ &
	$\mathcal{O}_{l^2 d^2 G}$ & $(\bar{l}\gamma^{\mu} l) (\bar{d}\gamma^{\nu} T^A d) G^A_{\mu\nu}$ \\
	
	$\mathcal{O}_{q^2 e^2 \tilde{G}}$ & $(\bar{e}\gamma^{\mu} e) (\bar{q}\gamma^{\nu} T^A q) \tilde{G}^A_{\mu\nu}$ &
	$\mathcal{O}_{q^2 e^2 G}$ & $(\bar{e}\gamma^{\mu} e) (\bar{q}\gamma^{\nu} T^A q) G^A_{\mu\nu}$ \\	
	
  \hline
  \end{tabular}
    \mycaption{Dimension-8 four-fermion operators with a gluon field that contribute to the Drell-Yan transverse momentum spectrum, organized according to their $CP$ transformation properties. $\tilde{G}$ denotes the dual field-strength tensor.  \label{tab:ffd8pT}}
\end{center}
\end{table}

\section{Example UV models} \label{sec:UV}

Our primary interest in this paper is the study of the SMEFT from the bottom-up, without reference to explicit UV models. Without 
further experimental guidance as to the form of new physics it is important to fully explore the possible parameter space without the introduction of theoretical biases. Another motivation of our work is to determine what particular experimental data sets, in this case the measurement of the Drell-Yan transverse momentum spectrum at high invariant mass, can teach us about different sectors of the SMEFT.
However, we do wish to demonstrate how the operators enumerated in Section~\ref{sec:smeft} can be 
obtained from explicit UV examples. In particular this study will show that the dimension-8 effects can serve as a useful diagnostic tool to distinguish between different UV states. We will study only example heavy particles here, and briefly 
mention how they can be embedded into full UV models. More detailed examples of the matching of full UV models to the dimension-8 level are discussed in the literature~\cite{Cohen:2020qvb,Dawson:2022cmu}. In our numerical studies we will focus on the example of all-singlet fermion operators. We consider two UV examples that can lead the operators of interest here: a $\zp$ and a vector leptoquark.

\subsection{Right-handed $\zp$ model}
\label{subsec:zprime}

We first study a $\zp$ boson coupled to SM singlets. The Lagrangian for this state is given by 
\be
{\cal L}_{\zp} = -\frac{1}{4}Z^{\prime \mu\nu} Z^{\prime}_{\mu\nu} +\frac{M_{\zp}^2}{2}-g_{\zp} \sum_f g_R^f\bar{\psi}_f \gamma^{\mu} P_R \psi_f \zp_{\mu}.
\ee
Here, $g_R^f$ denotes the charge of fermion $f$ under the U(1) gauge group, while $g_{\zp}$ is an overall coupling strength of the $\zp$ that is extracted from the charges. Although it is not our intent here to discuss full UV models we note that this state can be embedded into an anomaly-free U(1) gauge theory with additional fermionic matter that can be taken heavy, in which cases the charges $g_R^f$ are fixed~\cite{Carena:2004xs}. To determine the Wilson coefficients this Lagrangian leads to for the operators introduced in Section~\ref{sec:smeft} we compute the process $u_1\bar{u}_2 \to l_3\bar{l}_4$, to fix $C_{eu}$ and $C^{(1)}_{e^2u^2D^2}$. We also compute $u_1\bar{u}_2 \to l_3\bar{l}_4g_5$ to fix $C_{e^2u^2\tilde{G}}$. A straightforward calculation of the amplitude expanded in the limit $s \ll M_{\zp}^2$, where $s=(p_1+p_2)^2$ is the usual partonic Mandelstam invariant, leads to 
\be
{\cal M}_{\zp}(u_1\bar{u}_2 \to l_3\bar{l}_4) = -g_{\zp}^2 g_R^q g_R^e \left\{ \frac{1}{M_{\zp}^2}+\frac{s}{M_{\zp}^4}+\ldots \right\} \bar{u}_3 \gamma^{\mu} P_R v_4 \bar{v}_2 \gamma_{\mu}P_R u_1.
\ee
From this we can read off the Wilson coefficients:
\bea
\frac{C_{eu}}{\Lambda^2} &=& -\frac{g_{\zp}^2 g_R^u g_R^e}{ M_{\zp}^2}, \nonumber \\
\frac{C^{(1)}_{e^2u^2D^2}}{\Lambda^4}  &=& -\frac{g_{\zp}^2 g_R^u g_R^e}{ M_{\zp}^4}.
\eea
An identical matching calculation for the down-quark channel fixes $C_{ed}$ and $C^{(1)}_{e^2d^2D^2}$. For simplicity, in the rest of this work we neglect these down-quark Wilson coefficients, and focus on the up-quark sector. We note that upon factoring out the dependence on the dimensionful scale $M_{\zp}$ the Wilson coefficients at dimension-6 and dimension-8 are identical in magnitude; there is no suppression of the dimension-8 coefficient.

We now consider the process $u_1\bar{u}_2 \to l_3\bar{l}_4g_5$. At tree-level two diagrams contribute, with the additional gluon radiated from either initial-state quark. In both cases there is no hard scale in the virtual quark propagator, and the only expansion possible is for the $\zp$ propagator. This indicates that this process is completely determined by the emission of a gluon after the insertion of either $\mathcal{O}_{eu}$ or $\mathcal{O}_{e^2u^2D^2}$, and consequently
\be
\frac{C_{e^2u^2\tilde{G}}}{\Lambda^4} =0.
\ee

\subsection{Vector leptoquark model}
\label{subsec:lepto}

We now consider a vector leptoquark coupled to right-handed leptons and quarks. The general Lagrangian for such a state is given in Refs.~\cite{Blumlein:1992ej,Blumlein:1996qp}. We assume a leptoquark coupled to QCD, and coupled to right-handed up quarks and leptons, for which the Lagrangian takes the form
\be
{\cal L}_{U} = -\frac{1}{2} G^{i\dagger}_{\mu\nu} G_i^{\mu\nu} +M_U^2 U_{\mu}^{i\dagger}U^{\mu}_i + h_U \left(\bar{\psi}_u^i \gamma^{\mu} P_R l \right)U^i_{\mu}-ig_s (1-\kappa_U) U^{\dagger}_{\mu}T^a U_{\nu} {\cal G}^{a\mu\nu}.
\ee
Here, the Roman indices $i,j$ denote color indices in the fundamental representation. The quantity ${\cal G}$ denotes the field strength tensor of the SM gluon field. The field strength tensor and covariant derivatives of the leptoquark are given by
\bea
D_{\mu}^{ik} &=& \partial_{\mu}\delta^{ij} -i g_s T_A^{ij} G^A_{\mu}, \nonumber \\
G^i_{\mu\nu} &=& D^{ik}_{\mu} U_{\nu k} - D^{ik}_{\nu} U_{\mu k}.
\eea
The coupling $\kappa_U$ is related to the magnetic moment of the leptoquark. We note that it has been argued that complete leptoquark models generically contain $\zp$ bosons as well~\cite{Baker:2019sli}. These lead to Wilson coefficient contributions similar to those discussed in Section~\ref{subsec:zprime}, and are not explicitly considered here since we focus instead on the unique aspects of the vector leptoquark. We also note that leptoquarks have received renewed interest recently due to their possible role in resolving outstanding flavor anomalies~\cite{Baker:2019sli}.

We compute the same partonic processes as before to determine the Wilson coefficients. A straightforward calculation of $u_1\bar{u}_2 \to l_3 \bar{l}_4$ leads, after a Fierz rearrangement, to the amplitude
\be
{\cal M}_{U} = \frac{h_U^2}{M_U^2} \bar{u}_3 \gamma^{\mu}P_R v_4 \bar{v}_2 \gamma_{\mu}P_R u_1\left\{ 1+\frac{t}{M_U^2} +\ldots \right\}.
\ee
In order to match this amplitude to the operators considered previously we have to decompose the $t$ in the numerator according to the contributions to the operators $\mathcal{O}^{(1)}_{e^2u^2D^2}$ and $\mathcal{O}^{(2)}_{e^2u^2D^2}$. Doing so we arrive at the Wilson coefficients for the operators considered here:
\bea
\frac{C_{eu}}{\Lambda^2} &=& \frac{h_U^2}{ M_{U}^2}, \nonumber \\
\frac{C^{(1)}_{e^2u^2D^2}}{\Lambda^4}  &=& -\frac{h_U^2}{ 4M_{U}^4}.
\eea
We note that the leptoquark also contributes to $C^{(2)}_{e^2u^2D^2}$, unlike the $\zp$. This is the first example of how dimension-8 coefficients can help distinguish between models. A non-zero $C^{(2)}_{e^2u^2D^2}$, which can be determined from an analysis of the type considered in~\cite{Alioli:2020kez}, would disfavor a $\zp$ UV model.

Another difference between the $\zp$ and leptoquark comes when we consider the gluonic process $u_1 \bar{u}_2 \to l_3 \bar{l}_4 g_5$. There is a tri-linear coupling $UUg$ which means that the gluon can be emitted from the $t$-channel leptoquark. Upon expanding around large $M_U$, this leads to a local contribution described by the operator $\mathcal{O}_{e^2 u^2 \tilde{G}}$. We relegate details of the matching calculation to  Appendix~\ref{sec:applepto}, and simply note the result here:
\be
\frac{C_{e^2 u^2 \tilde{G}}}{\Lambda^4} = -\frac{h_U^2 g_s(1-\kappa_U)}{2 M_U^4}
\ee
This again illustrates our point that measurement of the complete sector of four-fermion operators can discriminate between UV models. $C_{e^2 u^2 \tilde{G}}$ is not induced in $\zp$ models, but it is in vector leptoquark models. We note as well the following points regarding the Wilson coefficients found in this calculation: upon removal of the dimensionful quantity $M_U$ the dimension-6 and dimension-8 Wilson coefficients are similar in size, and the coupling $C_{e^2 u^2 \tilde{G}}$ can be larger than $C^{(1)}_{e^2u^2D^2}$ for negative $\kappa_U$. We will refer to both of these points during discussions in later sections. It has been pointed out that positivity bounds on dimension-8 Wilson coefficients can be derived from the underlying principles of quantum field theory~\cite{Zhang:2020jyn,deRham:2022hpx}. We note that the simplest elastic positivity constraints do not restrict the parameter space of the $C^{(1)}_{e^2u^2D^2}$ and $C_{e^2 u^2 \tilde{G}}$ coefficients~\cite{Li:2022rag}.

\section{Motivation for the doubly-differential Drell-Yan distribution} \label{sec:doubdiff}

\begin{figure}[h!]
\centering
\includegraphics[width=0.46\textwidth]{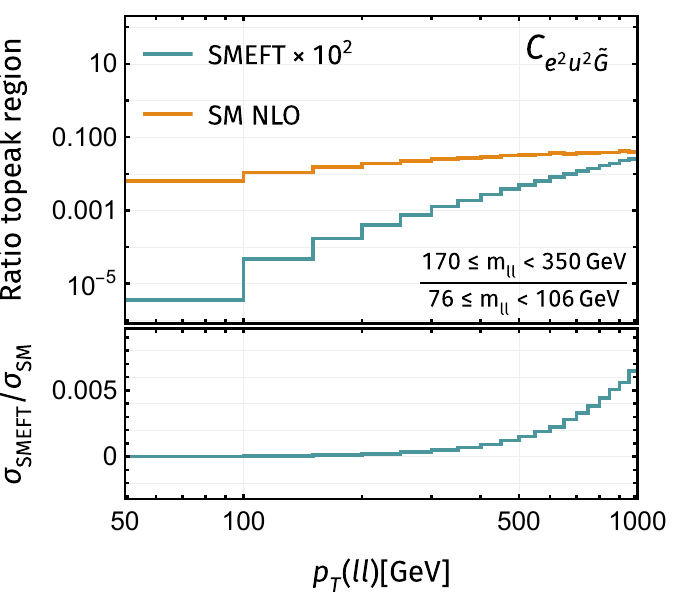}
\includegraphics[width=0.46\textwidth]{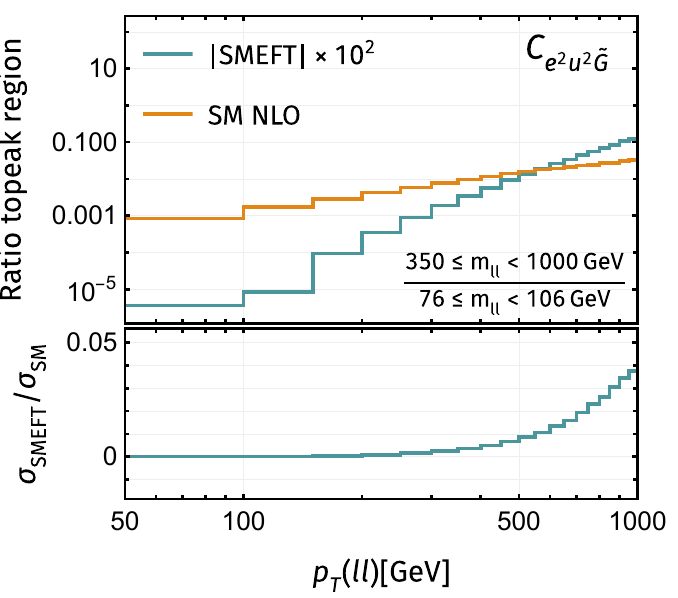}
\caption{Ratio of the SMEFT correction assuming non-zero $C_{e^2u^2\tilde{G}}$ to the SM result as a function of $p_T$ for the upper two invariant mass bins of the CMS measurement~\cite{CMS:2022ubq}. The results have been normalized to the $Z$-peak region. 
\label{fig:Ce2u2Gratio}}
\end{figure}

A main motivation of our work is to demonstrate that LHC data sets not originally intended as new physics searches can be sensitive, sometimes in novel ways, to the SMEFT parameter space, and therefore have unexpected sensitivity to physics beyond the SM. As an example, the CMS experiment has measured the lepton-pair transverse momentum spectrum in several invariant bins ranging up to 1 TeV~\cite{CMS:2022ubq}. Results are normalized to the $Z$-peak region in order to reduce the dependence of the measurement on systematic errors. The measurement was performed at 13 TeV with 36.3 fb$^{-1}$ of integrated luminosity. Before performing 
fits of this data to the SMEFT framework to demonstrate its sensitivity we describe why this data set is particularly interesting to probe the set of operators described in the previous sections. We show below in Figs.~\ref{fig:Ceuratio}~\ref{fig:Ce2u2D2ratio}, and~\ref{fig:Ce2u2Gratio} the ratio of SMEFT corrections to the SM result as a function of transverse momentum for the two highest invariant mass bins available in the measurement of Ref.~\cite{CMS:2022ubq}, turning on the three Wilson coefficients $C_{eu}$, $C_{e^2u^2D^2}$, and $C_{e^2u^2\tilde{G}}$ separately (we drop the superscript on $C_{e^2u^2D^2}$ henceforth, since we do not consider $C_{e^2u^2D^2}^{(2)}$ further in this work). The SM has been computed at NLO in QCD using the program MCFM~\cite{Campbell:2019dru}. We have set each Wilson coefficient to unity when making these plots. Since the coefficient $C_{eu}$ contributes first at dimension-6 it shows the largest deviations from the SM result. However, the deviation does not vary significantly with $p_T$. This is consistent with the structure of the dimension-6 EFT correction, which does not depend on the momentum flow into the effective vertex. The deviations for both $C_{e^2u^2D^2}$ and $C_{e^2u^2\tilde{G}}$ increase with $p_T$, consistent with the fact that the effective vertex depends on the momentum flow. While the $C_{e^2u^2D^2}$ deviation increases moderately with transverse momentum, $C_{e^2u^2\tilde{G}}$ increases rapidly. The EFT vertex for this operator is directly proportional to the gluon momentum, and therefore the $p_T$ of the lepton pair by momentum conservation. This is the motivation for the analysis of this data set within the SMEFT framework: the $p_T$ distribution offers additional sensitivity to gluonic operators not present with invariant 
mass distributions alone.

\begin{figure}[h!]
\centering
\includegraphics[width=0.46\textwidth]{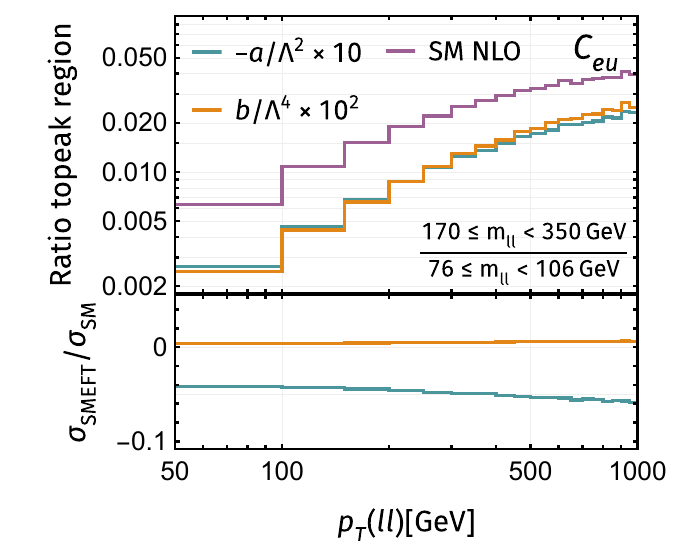}
\includegraphics[width=0.46\textwidth]{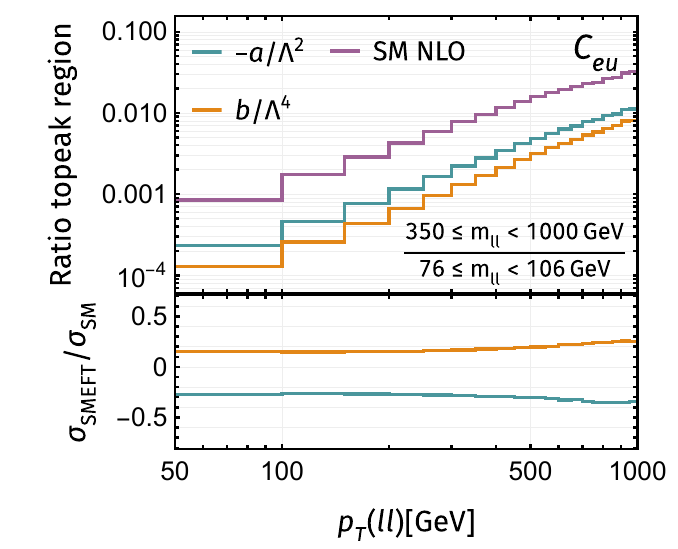}
\caption{Ratio of the SMEFT correction assuming non-zero $C_{eu}$ to the SM result as a function of $p_T$ for the upper two invariant mass bins of the CMS measurement~\cite{CMS:2022ubq}. The results have been normalized to the $Z$-peak region. The $a$-term corresponds to the linear $1/\Lambda^2$ correction while the $b$-term corresponds to the quadratic  $1/\Lambda^4$ correction. 
\label{fig:Ceuratio}}
\end{figure}

\begin{figure}[h!]
\centering
\includegraphics[width=0.46\textwidth]{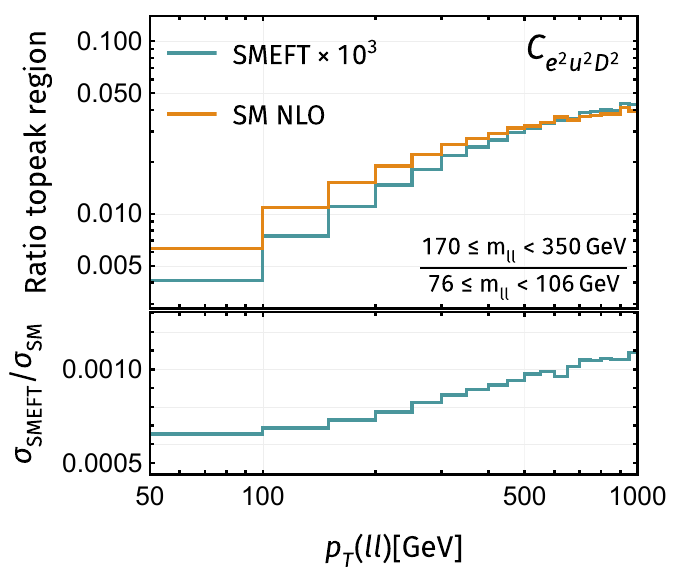}
\includegraphics[width=0.46\textwidth]{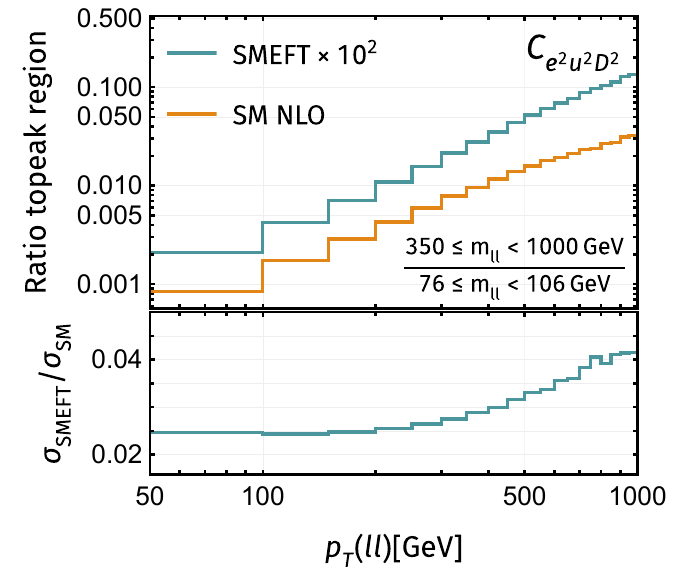}
\caption{Ratio of the SMEFT correction assuming non-zero $C_{e^2u^2D^2}$ to the SM result as a function of $p_T$ for the upper two invariant mass bins of the CMS measurement~\cite{CMS:2022ubq}. The results have been normalized to the $Z$-peak region. 
\label{fig:Ce2u2D2ratio}}
\end{figure}

\section{Calculational framework and fits to the current data}
\label{sec:current}

We begin by performing a fit of the current CMS measurement to the SMEFT framework. Although we will find that there is limited sensitivity to the gluonic operators at this point, we find it useful to quantify the current sensitivity and to also establish our notation for later sections. We will for simplicity focus on operators that contain right-handed fermions only. Possible UV models leading to such operators were discussed in Section~\ref{sec:UV}, where we introduced the possibility of using $C_{e^2u^2\tilde{G}}$ to distinguish between them. We focus on invariant mass bins above the $Z$-peak, since the SMEFT-induced corrections grow with energy. We also focus on transverse momentum bins above 50 GeV, following the same logic as for invariant mass, and also to avoid phase space regions where $p_T$ resummation may play a role. This leaves us with the eight bins shown below in Table~\ref{tab:CMSbins}. We note that these bins are normalized by the experiment to the $Z$-peak region $76 \leq m_{ll} \leq 106$ GeV. 
\begin{table}[h!]
\centering
\begin{tabular}{|c|c|}
\hline
$m_{ll}=[106,170]$ GeV & $p_T=[52,85,160,250]$ GeV  \\
$m_{ll}=[170,350]$ GeV & $p_T=[52,160]$ GeV  \\	
$m_{ll}=[350,1000]$ GeV & $p_T=[52,160]$ GeV  \\		
  \hline
\end{tabular}
\mycaption{Bins in invariant mass and transverse momentum used in Ref.\cite{CMS:2022ubq}. The $p_T$ values refer to bin boundaries. There is an upper cut of 1 TeV on $p_T$ in all bins. \label{tab:CMSbins}}
\end{table}

We now give the details of our calculational framework. We compute the SM cross section at next-to-leading order (NLO) in QCD using the MCFM program~\cite{Campbell:2019dru}. We use the NNPDF 3.1 NLO parton distribution functions~\cite{NNPDF:2017mvq}.  To compute the PDF errors we follow the standard procedure for Monte Carlo replica sets~\cite{NNPDF:2017mvq}. To estimate the error arising from higher-order QCD corrections we set the renormalization and factorization scales to the central value
\be
\mu_0=\sqrt{m_{ll}^2+p_T(ll)^2}
\ee
and vary them around this value in an uncorrelated way according to 
\be
\frac{1}{2} \leq \mu_{R,F}/\mu_0 \leq 2, \;\; \frac{1}{2} \leq \mu_{R}/\mu_F \leq 2.
\ee
We find the largest variation within this range, and form a symmetric scale uncertainty using this largest variation. We note that this technique leads to slightly more conservative errors than the usual approach, in which the width of the scale variation band without symmetrization is used. We note that the PDF uncertainties are strongly correlated between different bins. We assume that the scale uncertainties are uncorrelated between bins. 

Since this data involves bins at high energies, it is important to quantify the effect of electroweak Sudakov corrections. We are unaware of a publicly available code that computes the electroweak Sudakov logarithms as a function of $p_T$ off the $Z$-peak. To estimate the impact of these corrections we compute the next-to-leading-logarithmic electroweak Sudakov corrections~\cite{Denner:2006jr} for each invariant mass bin integrated inclusively over $p_T$, apply this correction to each of the $p_T$ bins for that invariant mass, and assign half of this correction as an additional theoretical error. The lower bin boundaries in $p_T$, which provide the largest contributions to each bin, do not go above 160 GeV. For the higher two invariant mass bins in our analysis this $p_T$ value is less than the invariant mass. Since the Sudakov logarithms grow with the Mandelstam variable $s$ that enters the process, and $s$ is dominated by the lepton invariant mass for the reason stated above, we believe that this is a reasonable estimate. We note that their effect ranges from 1\% to 4\% as we increase the invariant mass bin, and has little effect on the quality of the fit to data. For the SMEFT cross section, we work at leading order in QCD.

We use the experimental uncertainties as provided by the CMS collaboration. These uncertainties range from $1.5\%$ to $8.9\%$, increasing with both invariant mass and transverse momentum, and contain a mix of both correlated and uncorrelated errors. The systematic uncertainties are dominant in the high invariant mass bins. We define a $\chi^2$ test to quantify the deviation of the SMEFT cross sections from the SM:
\begin{align}
    \chi^2=\sum_{i,j}^{\textrm{\#of bins}}\frac{\left(\sigma_{i}^{\mathrm{SM}}-\sigma_{i}^{\mathrm{SMEFT}}\right)\left(\sigma_{j}^{\mathrm{SM}}-\sigma_{j}^{\mathrm{SMEFT}}\right)}{\Delta \sigma_{i j}^{2}},
\end{align}
where $\Delta\sigma^2_{ij}$ signifies the error matrix composed of both theoretical and experimental uncertainties. We then extract the 95\% CL bounds of the Wilson coefficients based on $\chi^2$ fits. Before studying the SMEFT we note that the SM furnishes an acceptable fit to the data, with a $\chi^2$ per degree of freedom of 1.4.

\begin{figure}[!hbtp]
    \centering
    \includegraphics[width=.45\linewidth]{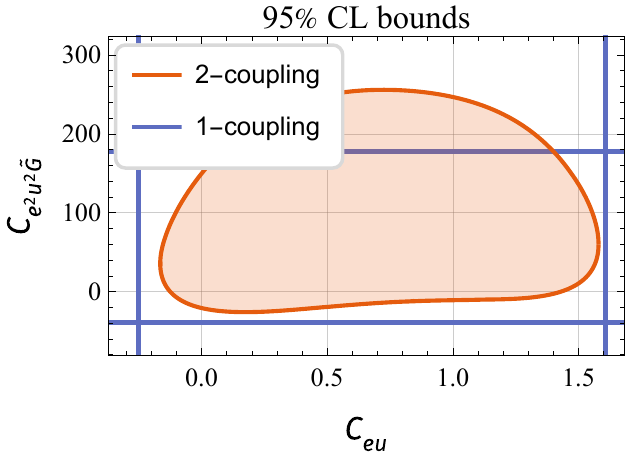}
    \includegraphics[width=.45\linewidth]{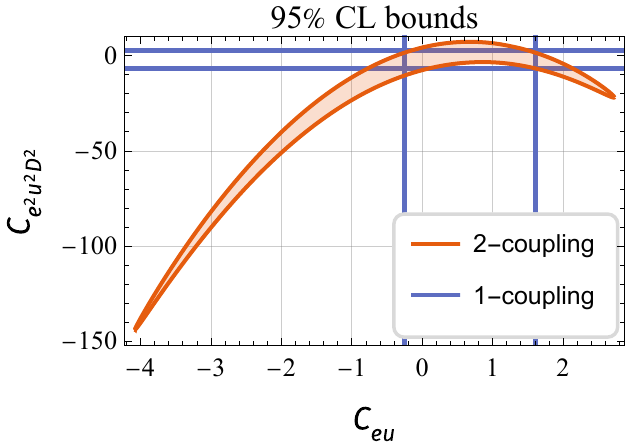}
    \caption{$95\%$ CL ellipses for the Wilson coefficients $C_{eu}$, $C_{e^2u^2\tilde G}$ and $C_{e^2u^2D^2}$ using current CMS data. For each diagram, one dimension-6 operator and one dimension-8 operator are enabled. The blue lines denotes the bounds with only one of the operators enabled. The energy scale $\Lambda$ is set to $1\;\mathrm{TeV}$. }
    \label{fig:CMS_2-d_fit_full}
\end{figure}

We consider turning on only single operator at a time, turning on pairs of operators, and turning on all three.
Figure~\ref{fig:CMS_2-d_fit_full} shows the $95\%$ CL ellipses of $C_{eu}$ together with either $C_{e^2u^2\tilde G}$ or $C_{e^2u^2D^2}$, as well as the bounds with only one Wilson coefficient enabled. Although the data is less sensitive to $C_{e^2u^2\tilde G}$ than to $C_{e^2u^2D^2}$, the circular nature of the ellipse in Figure~\ref{fig:CMS_2-d_fit_full} indicates that $C_{e^2u^2\tilde G}$ has little correlation with $C_{eu}$. The stretched narrow ellipse of $C_{eu}$ and $C_{e^2u^2D^2}$ shows a strong correlation between these two operators, with the effects of the two coefficients indistinguishable with the current data.

A potential issue that must be addressed when studying these constraints is the convergence of the EFT expansion. As can be seen from the examples in Section~\ref{sec:UV} we have the following rough relations between parameters in UV models and those appearing the SMEFT:
\bea
\frac{C}{\Lambda^2} &\sim & \frac{g^2}{M^2} \;\; \text{for dimension-6}, \nonumber \\
\frac{C}{\Lambda^4} &\sim & \frac{g^2}{M^4} \;\; \text{for dimension-8}.
\label{eq:masseffscale}
\eea
We want any potential resonance to lie above the scales probed experimentally. We take this constraint to be $M > 1$ TeV, the highest scale probed in this data, which translates to the bounds
\bea
\frac{\Lambda}{\sqrt{|C|}} > \frac{1 \;\text{TeV}}{g} \;\;\; \text{for dimension-6}, \nonumber \\
\frac{\Lambda}{\sqrt[4]{|C|}} > \frac{1\;\text{TeV}}{\sqrt{g}} \;\;\; \text{for dimension-8}.
\eea
The exact numerical value of this constraint depends on the coupling $g$, and therefore on the details of the UV model. We take the strong coupling limit $g=\sqrt{4\pi}$, leading to the least stringent constraint, in order to avoid ruling out allowed parameter space. This slightly restricts the allowed parameter space in the joint $C_{eu}, C_{e^2u^2\tilde G}$ in Fig.~\ref{fig:CMS_2-d_fit_full}. It does not affect the joint $C_{eu}, C_{e^2u^2D^2}$ result.

We list the $95\%$ CL bounds with only one operator enabled in Table~\ref{tab:CMSbounds}. We also calculate the bounds with multiple operators by marginalizing over the couplings. We further impose the effective scale constraint on the marginalized bounds in the final column of this table. The results are also listed in Table~\ref{tab:CMSbounds}. We observe that $C_{e^2u^2D^2}$ has a strong impact to the bounds on $C_{eu}$, while $C_{e^2u^2\tilde G}$ only mildly changes this limit. Turning on $C_{e^2u^2D^2}$ significantly weakens the bounds on $C_{eu}$, an effect also observed with LHC invariant mass distributions in~\cite{Boughezal:2021tih}. In general there is limited sensitivity of this data set to the dimension-8 coefficients with Wilson coefficients reaching ${\cal O}(100)$ still allowed.
\begin{table}[!h]
    \centering
	\begin{tabular}{|c||c||c|c|}
    	\hline
    	Wilson coefficient&Single coupling&Marginalized&Marginalized\textsuperscript{*}\\
    	\hline
    	\multicolumn{4}{|c|}{$C_{eu}\&C_{e^2u^2D^2}$}\\\hline
    	$C_{eu}$&${{[-0.252, 1.61]}}$&$[-3.44,2.37]$&${{[-3.34,2.33]}}$\\
    	$C_{e^2u^2D^2}^{(1)}$ 
    	&$[-6.65, 2.64]$&$[-90.9,47.1]$&$[-83.7,41.9]$\\
    	\hline
    	\multicolumn{4}{|c|}{$C_{eu}\&C_{e^2u^2\tilde{G}}$}\\\hline
    	$C_{eu}$&${{[-0.252, 1.61]}}$&$[-0.062,1.45]$&${{[-0.071,1.45]}}$\\
    	$C_{e^2u^2\tilde{G}}$&$[-39.4, 177.]$&$[-16.9,220]$&$[-8.95,197]$\\\hline
    	\multicolumn{4}{|c|}{all 3 operators}\\\hline
    	$C_{eu}$
    	&${{[-0.252, 1.61]}}$&$[-3.75,2.60]$&${{[-3.49,2.46]}}$\\
    	$C_{e^2u^2D^2}$
    	&$[-6.65, 2.64]$&$[-113.,57.2]$&$[-92.7,42.9]$\\
    	$C_{e^2u^2\tilde{G}}$ 
    	&$[-39.4, 177]$&$[-6.35,251.]$&$[3.94,201.0]$\\
    	\hline
    \end{tabular}
    \caption{$95\%$ CL bounds for the Wilson coefficients $C_{eu}$, $C_{e^2u^2\tilde G}$ and $C_{e^2u^2D^2}$ from current CMS data. The first column shows the bounds assuming one operator is enabled at a time. The second column shows the bounds on a given coefficient with the other enabled operators allowed to vary as well. The third column shows these bounds with the dimension-8 coefficients restricted according to the discussion in the main text. }
    \label{tab:CMSbounds}
\end{table}

\section{Fits to simulated HL-LHC data}
\label{sec:hllhc}

We found in the previous section that the current data shows little sensitivity to $C_{e^2u^2\tilde{G}}$. We consider next the potential the high-luminosity LHC to probe this SMEFT parameter space through a similar analysis. Since no data is yet available we resort to pseudodata generated with the NLO SM cross section. The HL-LHC pseudodata is generated under similar conditions as the CMS measurement~\cite{CMS:2022ubq}. We assume the center-of-mass energy $\sqrt{s}=14\;\textrm{TeV}$, an integrated luminosity $3\;\textrm{ab}^{-1}$ and a dilepton transverse momentum cut $p_T \geq 100\;\textrm{GeV}$. Since the eventual HL-LHC binning is unknown, we consider two possible sets of bins in the dilepton invariant mass and transverse momentum. The binning for the dilepton invariant mass $m_{ll}$ is motivated by the simulation in~\cite{Panico:2021vav}. For the binning of the dilepton transverse momentum $p_T(ll)$, we enforce that the relative statistical uncertainty of each bin cannot exceed $10\%$. As such, we discard the highest $m_{ll}$ bin in~\cite{Panico:2021vav} where $2600\leq m_{ll}\leq 14000\;\textrm{GeV}$. Next, two different binning strategies are applied: a coarse binning where the relative statistical uncertainty of each bin should be smaller than $5\%$ if possible~\footnote{The only exception is when $2000\leq m_{ll}\leq 2600\;\textrm{GeV}$, where the largest possible $p_T$ bin is $100\leq p_T\leq7000\;\textrm{GeV}$. The relative statistical uncertainty of this bin is larger than $5\%$, but still smaller than $10\%$. }; a fine binning where the relative statistical uncertainty of each bin must be smaller than $10\%$. 
We show the explicit bins used in Appendix~\ref{sec:appbinning}, with the coarse binning shown in Table~\ref{tab:bins_5pct} and the fine binning in Table~\ref{tab:bins_10pct}. 

We assume that the pseudodata is affected by three sources of experimental uncertainties: the statistical uncertainty $\Delta\sigma_{\textrm{stat}}$, the uncorrelated systematic uncertainty $\Delta\sigma_{\textrm{uncorr}}$ and the fully correlated systematic uncertainty $\Delta\sigma_{\textrm{corr}}$. We construct the cross section of bin $b$ using
\begin{align}
    \sigma_b^{\textrm{pseudo}}=\sigma_b^{\textrm{SM}}+r_b\sqrt{\Delta\sigma_{\textrm{stat},b}^2+\Delta\sigma_{\textrm{uncorr},b}^2}+r'\sqrt{\Delta\sigma_{\textrm{corr},b}^2},
\end{align}
where $r_b$ and $r'$ are random numbers generated with a normal distribution of mean $0$ and standard deviation $1$. For uncorrelated uncertainties, a separate random number is chosen for each bin.  For the correlated uncertainty, a single random number $r'$ is used across all bins. We assume the relative uncorrelated systematic uncertainty $\Delta\sigma_{\textrm{uncorr},b}/\sigma_b^{\textrm{SM}}=1\%$ and the relative correlated systematic uncertainty $\Delta\sigma_{\textrm{corr},b}/\sigma_b^{\textrm{SM}}=2\%$. These choices are consistent with the current values found in the CMS measurement~\cite{CMS:2022ubq}. We normalize the cross section to the Z-peak region $76\leq m_{ll}\leq 106\;\textrm{GeV}$. We emphasize that the relative uncorrelated uncertainty of the Z-peak region is still assumed as $1\%$, but any correlated uncertainties in this region are disregarded. The error matrix contains the experimental uncertainties, the theoretical PDF uncertainties and the scale uncertainties. The experimental part is constructed with $\Delta\sigma_{\textrm{stat}}$, $\Delta\sigma_{\textrm{uncorr}}$ and $\Delta\sigma_{\textrm{corr}}$. The PDF and scale uncertainties only contain SM contributions. For each set of random numbers $r_b$ and $r'$ generated, we perform a $\chi^2$ fit as described in Section~\ref{sec:current}. Each set of random numbers signifies one pseudo-experiment, and for each pseudo-experiment $e$, a set of best-fit Wilson coefficients $\{{C}_{i,e}\}$ is obtained. A total number of 1000 pseudo-experiments are evaluated, and the average best-fit values are obtained by
\begin{align}
\pmqty{\bar{C}_{1} \\
\bar{C}_{2}\\\vdots}=\left[\sum_{e=1}^{N_{\exp }}\left(V^{-1}\right)_{e}\right]^{-1}\left[\sum_{e=1}^{N_{\exp }}\left(V^{-1}\right)_{e}\pmqty{\bar{C}_{1, e} \\\bar{C}_{2, e}\\\vdots}\right].
\end{align}
The covariance matrix for each pseudo-experiment is given by
\begin{align}
\left(V^{-1}\right)_{i j}=\frac{1}{2} \frac{\partial \chi^{2}}{\partial C_{i} \partial C_{j}}. 
\end{align}
The $95\%$ CL bounds on the Wilson coefficients are extracted by
\begin{align}
    \pmqty{C_1-\bar{C}_1\\C_2-\bar{C}_2\\\vdots\\C_N-\bar{C}_N}^TV^{-1}\pmqty{C_1-\bar{C}_1\\C_2-\bar{C}_2\\\vdots\\C_N-\bar{C}_N}=\Delta\chi^2,
\end{align}
where $N$ is the number of Wilson coefficients. For $N=1,\,2,\,3$, $\Delta\chi^2=3.841,\,5.991,\,7.815$, respectively. The average inverse covariance matrix is defined as 
\begin{align}
V^{-1}=\frac{1}{N_{\exp }} \sum_{e=1}^{N_{\exp }}\left(V^{-1}\right)_{e}. 
\end{align}

\begin{figure}[!hbtp]
    \centering
    \includegraphics[width=.45\linewidth]{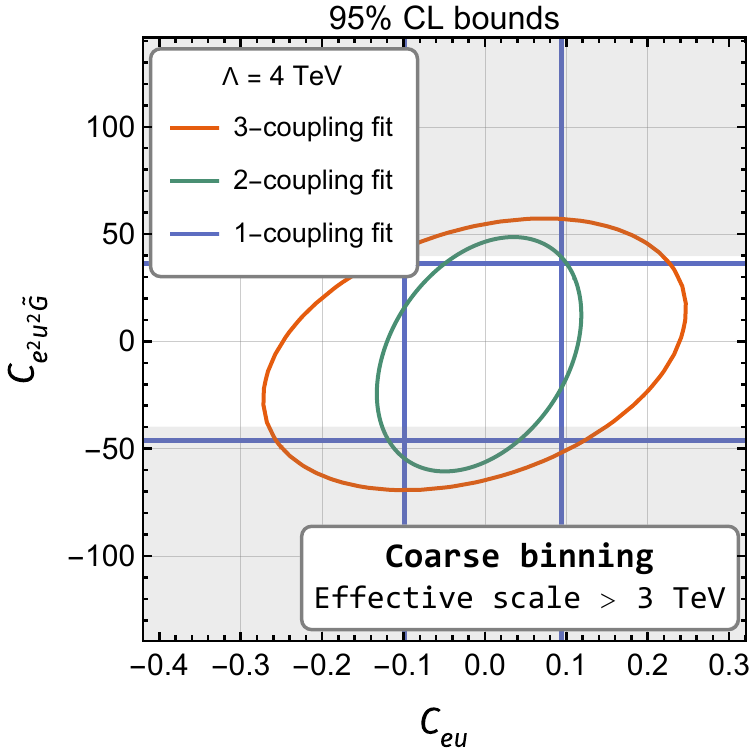}
    \includegraphics[width=.45\linewidth]{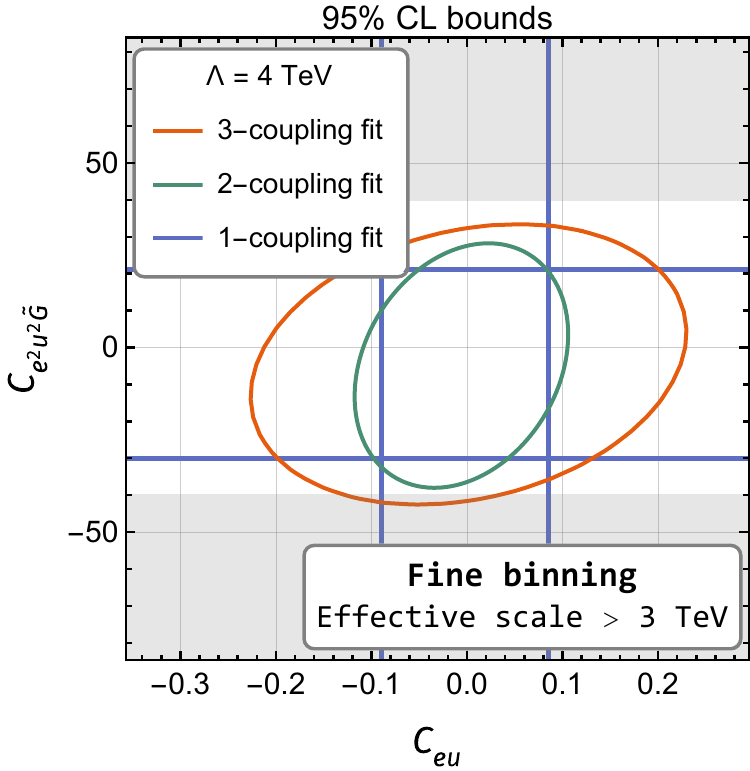}
    \caption{$95\%$ CL ellipses for the Wilson coefficients $C_{eu}$ and $C_{e^2u^2\tilde G}$ at the HL-LHC. The left diagram shows the bounds with the coarse binning, while the right one shows the bounds with the fine binning. The blue lines denote the bounds with only one of the operators enabled, the green line denotes the bounds with two operators enabled, and the orange line denotes the bounds with $C_{e^2u^2D^2}$ also enabled. The energy scale $\Lambda$ is set to $4\;\mathrm{TeV}$, and the effective scale constraint is set to $3\;\mathrm{TeV}$. The dark gray area shows the effective scale constraint with $g=\sqrt{4\pi}$.  }
    \label{fig:HL-LHC-G}
\end{figure}

\begin{figure}[!hbtp]
    \centering
    \includegraphics[width=.45\linewidth]{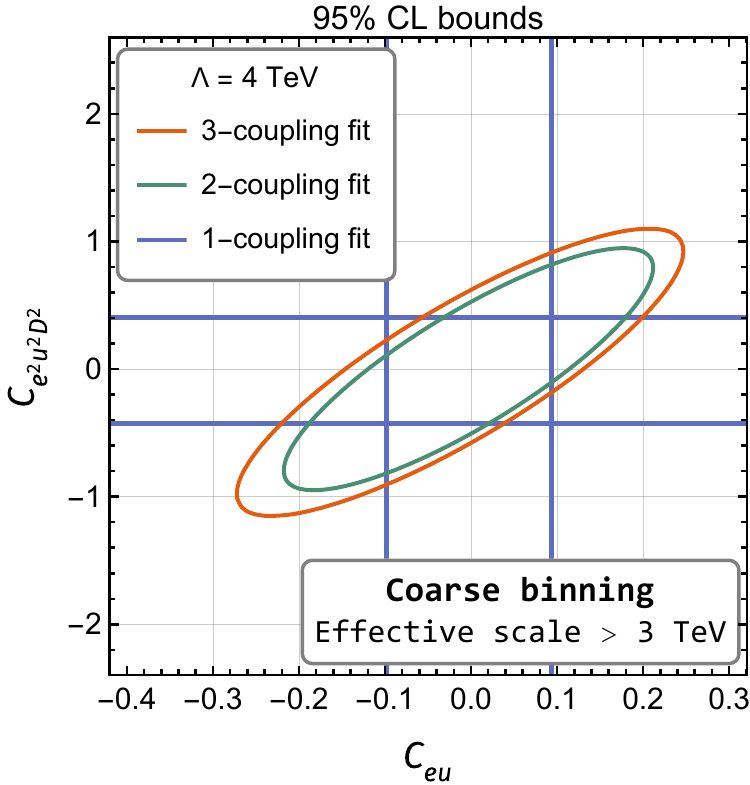}
    \includegraphics[width=.45\linewidth]{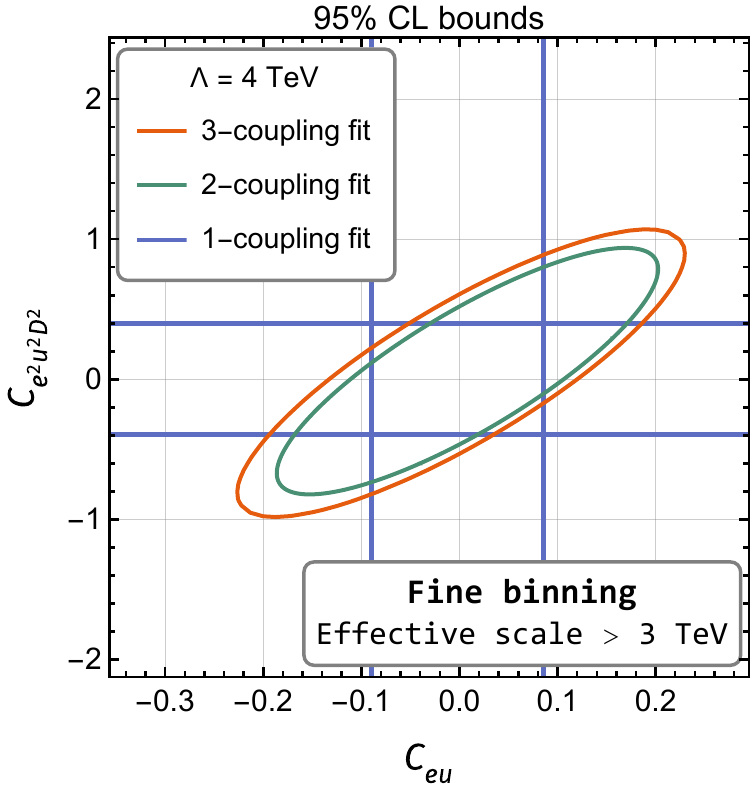}
    \caption{$95\%$ CL ellipses for the Wilson coefficients $C_{eu}$ and $C_{e^2u^2D^2}$ at the HL-LHC. The left diagram shows the bounds with the coarse binning, while the right one shows the bounds with the fine binning. The blue lines denotes the bounds with only one of the operators enabled, the green line denotes the bounds with two operators enabled, and the orange line denotes the bounds with $C_{e^2u^2\tilde G}$ also enabled. The energy scale $\Lambda$ is set to $4\;\mathrm{TeV}$, and the effective scale constraint is set to $3\;\mathrm{TeV}$. }
    \label{fig:HL-LHC-D}
\end{figure}

We now consider fits to the pseudodata with one, two or three Wilson coefficients enabled. Figure~\ref{fig:HL-LHC-G} shows the $95\%$ CL bounds on $C_{eu}$ and $C_{e^2u^2\tilde G}$ with either one, two or all three operators enabled. Figure~\ref{fig:HL-LHC-D} shows the $95\%$ CL bounds on $C_{eu}$ and $C_{e^2u^2D^2}$. The round shape of the ellipse in Figure~\ref{fig:HL-LHC-G} and the narrow shape in Figure~\ref{fig:HL-LHC-D} confirms what we learned from the existing CMS data: there is little correlation between $C_{eu}$ and $C_{e^2u^2\tilde G}$, and a stronger correlation between $C_{eu}$ and $C_{e^2u^2D^2}$. We observe that the inclusion of $C_{e^2u^2D^2}$  loosens the bounds on $C_{eu}$, while the inclusion of $C_{e^2u^2\tilde G}$ has much less impact on $C_{eu}$. This is consistent with the previous observation that the inclusion of transverse momentum data provides a separate handle on the gluonic operators. We also observe that the fine binning leads to tighter bounds than the coarse binning. The gray area in Figure~\ref{fig:HL-LHC-G} shows the region where the effective scale constraint of the dimension-8 operator is violated, and the EFT expansion is no longer valid. We demand the constraint $M>3\;\textrm{TeV}$, consistent with the upper limit of our invariant mass binning, so that the dimension-8 Wilson coefficients must satisfy $\frac{\Lambda}{\sqrt[4]{|C|}}>\frac{3 \mathrm{TeV}}{\sqrt{g}}$. We use dark gray to indicate $g=\sqrt{4\pi}$, a choice discussed in the previous section.

\begin{figure}[!hbtp]
    \centering
    \includegraphics[width=.85\linewidth]{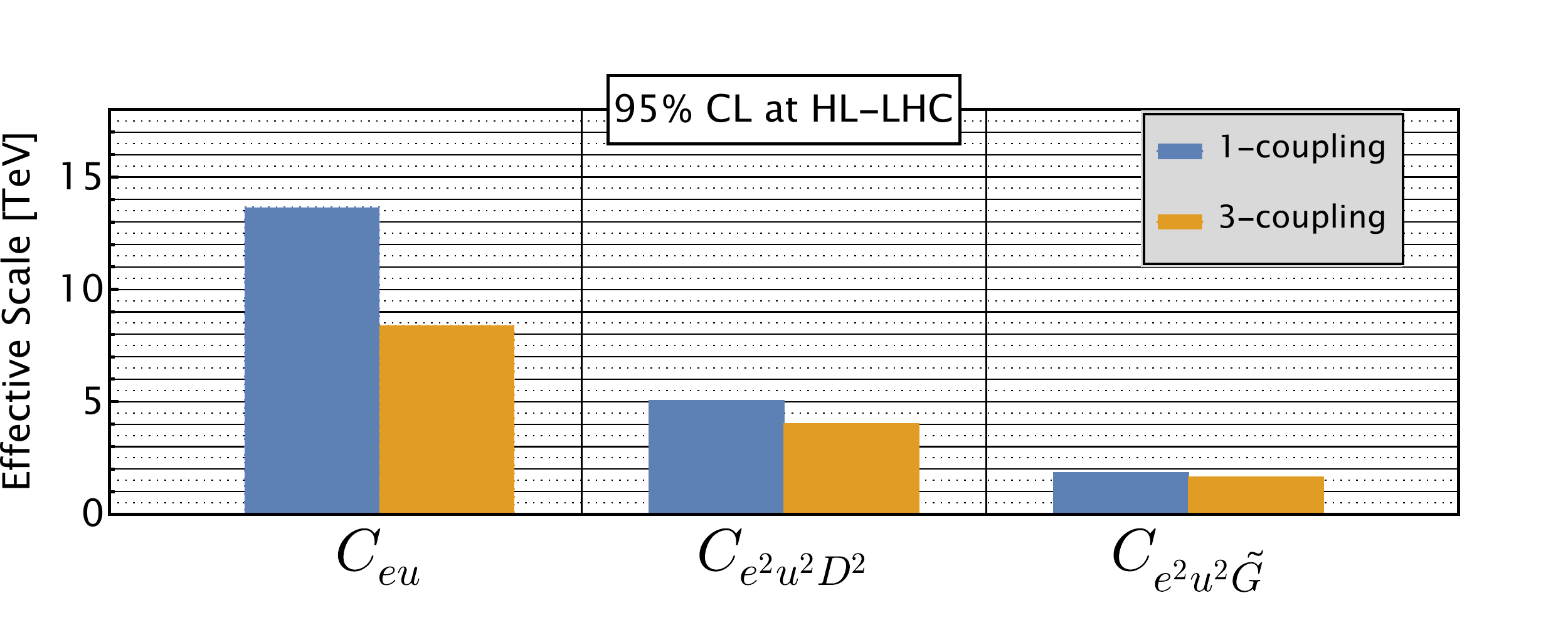}
    \caption{$95\%$ CL bounds on the effective scale of all three operators when only a single one is enabled, and when all three are simultaneously enabled, at the HL-LHC. We recall that the effective scale is defined as $\Lambda/\sqrt{C}$ for dimension-6 and $\Lambda/\sqrt[4]{C}$ for dimension-8. }
    \label{fig:effscale}
\end{figure}

We observe that the HL-LHC data has the potential to measure these three couplings separately. Although some correlation between $C_{eu}$ and $C_{e^2u^2D^2}$ remains, it is weaker than found with the current data, and only weakens the $C_{eu}$ bounds by a factor of two. Referring to the fine binning results, we observe that there is a hierarchy in the sensitivities to these three coefficients: $C_{eu}$ values of ${\cal O}(0.1)$ can be probed, $C_{e^2u^2D^2}$ values of ${\cal O}(1)$ can be probed, while the sensitivity to $C_{e^2u^2\tilde G}$ drops to ${\cal O}(10)$. Considering that the EFT expansion parameter is chosen as $\Lambda=4$ TeV, these results indicate sensitivity reaching into the multi-TeV region for all three operators. We recall that $C_{e^2u^2\tilde G}$ can be enhanced in certain regions of leptoquark parameter space, as discussed in Section~\ref{subsec:lepto}. This indicates that the HL-LHC doubly-differential Drell-Yan data can serve as a useful diagnostic tool for realistic UV states. We summarize the potential of the EIC by showing in Fig.~\ref{fig:effscale} the effective UV scales that can be probed for each parameter when only a single coupling is turned on, and when all three are turned on. We recall from Eq.~(\ref{eq:masseffscale}) that the effective scale is related to the heavy resonance mass in the UV theory scaled by either $g$ (for dimension-6) or $\sqrt{g}$ (for dimension-8) as shown in Section~\ref{sec:current}. Effective scales approaching 10 TeV can be probed for $C_{eu}$, while sensitivities reaching several TeV are possible for the dimension-8 coefficients.

\section{Conclusions}  \label{sec:conc}

In this paper we have studied probes of the semi-leptonic four-fermion sector of the SMEFT that are possible with neutral-current Drell-Yan measurements at the LHC. We have extended previous studies by including dimension-8 operators with additional gluon field-strength tensors. These operators directly modify the high transverse momentum region in Drell-Yan production. A motivation for this work is a recent CMS measurement of the transverse momentum distribution for the Drell-Yan process further binned in invariant mass. Although this work was intended primarily as a QCD study, it has novel BSM sensitivity as well, and provides direct access to this previously unexplored sector of the SMEFT. 

To motivate our study we have demonstrated that example UV models can lead to very different patterns of Wilson coefficients for these gluonic operators; some states generate potentially sizable Wilson coefficients for this dimension-8 operator, while others do not generate these operators. This ability to discriminate between different UV completions of the SMEFT would be missed if the SMEFT expansion was truncated at the dimension-6 level; the example models considered here would only match to a single dimension-6 operator. Measurement of the entire suite of semi-leptonic four-fermion coefficients through dimension-8 can therefore help distinguish between different models of new physics. We have considered fits of the SMEFT framework to both the current CMS measurement and to simulated future HL-LHC data. While the current data shows little sensitivity to the gluonic operator, there are good prospects for probing this effect with future data. We encourage  this measurement to be performed with future data and its BSM potential  to be further explored.

\section*{Acknowledgments}
We thank E.~Mereghetti for helpful comments. R.~B. is supported by the DOE contract DE-AC02-06CH11357.  Y.~H and F.~P. are supported
by the DOE grants DE-FG02-91ER40684 and DE-AC02-06CH11357.

\appendix

\section{Leptoquark matching}
\label{sec:applepto}

In this Appendix we study the matching of a vector leptoquark to the SMEFT in the process $u_1 \bar{u}_2 \to l_3 \bar{l}_4 g_5$ in order to determine the Wilson coefficient $C_{e^2 u^2 \tilde{G}}$. There are three contributing diagrams, two where the gluon is emitted from an initial quark, and one where it is emitted from the leptoquark. The amplitudes for each diagram take the form
\bea
i {\cal M}_1 &=& \frac{i h_U^2 g_s T^a_{ij}}{t_{15} (t_{24}-M_U^2)} \left\{ \bar{v}_2^i \gamma_{\mu}P_Rv_4 \bar{u}_3 \gamma^{\mu} P_Ru_1^j \frac{2p_1 \cdot \epsilon_5^a}{t_{15}} -\bar{v}_2^i \gamma_{\mu}P_Rv_4 \bar{u}_3 \gamma^{\mu}\dslash{p}_5 \dslash{\epsilon}_5^a P_Ru_1^j \right\}
	\nonumber \\
i {\cal M}_2 &=& -\frac{i h_U^2 g_s T^a_{ij}}{t_{25} (t_{13}-M_U^2)} \left\{ \bar{v}_2^i \gamma_{\mu}P_Rv_4 \bar{u}_3 \gamma^{\mu} P_Ru_1^j \frac{2p_2 \cdot \epsilon_5^a}{t_{25}} -\bar{v}_2^i \gamma^{\mu} \dslash{\epsilon}_5^a \dslash{p}_5 P_Rv_4 \bar{u}_3 \gamma_{\mu} P_Ru_1^j \right\}	
	\nonumber \\
i {\cal M}_3 &=& \frac{i h_U^2 g_s T^a_{ij}}{(t_{13}-M_U^2)(t_{24}-M_U^2)} \left\{ \bar{v}_2^i \gamma_{\mu}P_Rv_4 \bar{u}_3 \gamma^{\mu} P_Ru_1^j (p_1-p_2-p_3+p_4)\cdot \epsilon_5^a  \right. \nonumber \\
&&\left. + (1-\kappa_U)\bar{v}_2^i \gamma^{\nu}P_Rv_4 \bar{u}_3 \gamma^{\mu}  P_Ru_1^j [p_{5\mu}\epsilon^a_{5\nu} - 
	p_{5\nu}\epsilon^a_{5\mu}] \right\}
\eea
using the Lagrangian presented in Section~\ref{sec:UV}. We note that $t_{15} = (p_1-p_5)^2$, etc. We can expand these expressions in the large $M_U$. The dimension-6 contribution comes from the first two diagrams; the $t_{15},t_{25}$ in the denominator make it clear that these match to the emission of a gluon off of a dimension-6 four fermion operator. We focus here on the expansion to dimension-8. The contribution from each diagram is
\bea
i {\cal M}^{(8)}_1 &=& -\frac{i h_U^2 g_s T^a_{ij}}{M_U^4} \left\{ \frac{2 p_1 \cdot \epsilon_5^a t_{24}}{t_{15}} \bar{v}_2^i \gamma_{\mu}P_Rv_4 \bar{u}_3 \gamma^{\mu} P_Ru_1^j - \frac{t_{24}}{t_{15}}\bar{v}_2^i \gamma_{\mu}P_Rv_4 \bar{u}_3 \gamma^{\mu}\dslash{p}_5 \dslash{\epsilon}_5^a P_Ru_1^j \right\}
	\nonumber \\
i {\cal M}^{(8)}_2 &=& \frac{i h_U^2 g_s T^a_{ij}}{M_U^4} \left\{ \frac{2 p_2 \cdot \epsilon_5^a t_{13}}{t_{25}} \bar{v}_2^i \gamma_{\mu}P_Rv_4 \bar{u}_3 \gamma^{\mu} P_Ru_1^j - \frac{t_{13}}{t_{25}}\bar{v}_2^i \gamma^{\mu} \dslash{\epsilon}_5^a \dslash{p}_5 P_Rv_4 \bar{u}_3 \gamma_{\mu} P_Ru_1^j  \right\}	
	\nonumber \\
i {\cal M}^{(8)}_3 &=& \frac{i h_U^2 g_s T^a_{ij}}{M_U^4} \left\{ \bar{v}_2^i \gamma_{\mu}P_Rv_4 \bar{u}_3 \gamma^{\mu} P_Ru_1^j (p_1-p_2-p_3+p_4)\cdot \epsilon_5^a  \right. \nonumber \\
&&\left. + (1-\kappa_U) \bar{v}_2^i \gamma^{\nu}P_Rv_4 \bar{u}_3 \gamma^{\mu}  P_Ru_1^j [p_{5\mu}\epsilon^a_{5\nu} - 
	p_{5\nu}\epsilon^a_{5\mu}] \right\}.	
\eea
It is clear that the first two diagrams come from emitting a gluon from dimension-8 four-fermion operators, and do not match to a local operator with a gluon. This can be seen from the $t_{15},t_{25}$ in the denominator. The same is true for the first term of ${\cal M}^{(8)}_3$. This can be determined most simply by demanding gauge invariance: the amplitude must vanish upon replacing $\epsilon_5 \to p_5$. The first two diagrams are not invariant themselves. Only upon adding the first term of diagram three is gauge invariance satisfied. The last term of ${\cal M}^{(8)}_3$ is separately gauge invariant.

This leaves the last term of $ {\cal M}^{(8)}_3$ to match to a local dimension-8 operator $q\bar{q} l\bar{l}g$. To simplify this we apply the 
following Fierz identity:
\be
 \bar{v}_2^i \gamma^{\nu}P_Rv_4 \bar{u}_3 \gamma^{\mu}  P_Ru_1^j = \frac{1}{2}\left\{ -g^{\mu\nu}g^{\rho\sigma} 
 	+g^{\mu\rho}g^{\nu\sigma}+g^{\mu\sigma}g^{\nu\rho}-i \epsilon^{\mu\nu\rho\sigma} \right\}
	\bar{v}_2^i \gamma^{\sigma}P_Ru_1^j  \bar{u}_3 \gamma^{\rho}  P_Rv_4.
	\label{eq:Fierz}
\ee
Only the antisymmetric term survives when we plug this into the amplitude, leaving us with
\be
i {\cal M}^{(8)}_{local} = \frac{h_U^2 (1-\kappa_U) g_s T^a_{ij}}{M_U^4}p_{5\mu}\epsilon_{5\nu}^a \epsilon^{\mu\nu\rho\sigma} \bar{v}_2^i \gamma^{\sigma}P_Ru_1^j  \bar{u}_3 \gamma^{\rho}  P_Rv_4.
\ee
This matches to the local dimension-8 operator
\be
\bar{e}\gamma^{\mu}e \bar{u}\gamma^{\nu}T^a u \tilde{G}^a_{\mu\nu}
\ee
with the Wilson coefficient:
\be
C_{e^2 u^2 \tilde{G}} = - \frac{h_U^2 (1-\kappa_U) g_s}{2 M_U^4}.
\ee

\section{HL-LHC binning}
\label{sec:appbinning}

We present in this Appendix the two choices for HL-LHC binning used in our analysis: a coarse binning where the relative statistical uncertainty of each bin is smaller than $5\%$, and a fine binning where the relative statistical uncertainty of each bin must be smaller than $10\%$.  The coarse binning is shown in Table~\ref{tab:bins_5pct}, while the fine binning is shown in Table~\ref{tab:bins_10pct}.

\begin{table}[!h]
    \centering
    \begin{tabular}{|c|c|}
		\hline
		$m_{ll}\;[\mathrm{GeV}]$&$p_T\;[\mathrm{GeV}$]\\
		\hline 
		$300-360$&\makecell{$[100, 110, 120, 130, 140, 150, 160, 170, 180, 190, 200, 210, 220, 230, 250, $\\$
		270, 290, 310, 330, 360, 380, 410, 440, 490, 570, 7000]$}\\\hline
		$360-450$&\makecell{$[100, 110, 120, 130, 140, 150, 160, 170, 180, 200, 230, 250, 270, 290, 310, $\\$
		330, 350, 370, 400, 440, 490, 580, 7000]$}\\\hline
		$450-600$&\makecell{$[100, 110, 120, 130, 140, 150, 160, 170, 180, 190, 210, 230, 250, 270, 290, $\\$
		320, 340, 360, 390, 430, 480, 580, 7000]$}\\\hline
		$600-800$&\makecell{$[100, 110, 120, 130, 150, 170, 200, 220, 250, 290, 320, 360, 420, 520, 7000]$}\\\hline
		$800-1100$&\makecell{$[100, 110, 120, 150, 170, 200, 230, 270, 330, 430, 7000]$}\\\hline
		$1100-1500$&\makecell{$[100, 200, 290, 7000]$}\\\hline
		$1500-2000$&\makecell{$[100, 7000]$}\\\hline
		$2000-2600$&\makecell{$[100, 7000]$}\\\hline
	\end{tabular}
    \caption{The coarse binning where the relative statistical uncertainty of each bin should be smaller than $5\%$ if possible. The first column shows the ranges of the $m_{ll}$ bins, and the second column shows the boundaries of the $p_T$ bins. }
    \label{tab:bins_5pct}
\end{table}

\begin{table}[!h]
    \centering
    \begin{tabular}{|c|c|}
		\hline
		$m_{ll}\;[\mathrm{GeV}]$&$p_T\;[\mathrm{GeV}]$\\
		\hline 
		$300-360$&\makecell{$[100, 110, 120, 130, 140, 150, 160, 170, 180, 190, 200, 210, 220, 230, 250, 270, $\\$
		290, 310, 330, 350, 370, 400, 420, 440, 470, 500, 530, 560, 600, 660, 760, 7000]$}\\\hline
		$360-450$&\makecell{$[100, 110, 120, 130, 140, 150, 160, 170, 180, 190, 200, 210, 220, 240, 260, 290, $\\$
		310, 330, 350, 370, 390, 410, 440, 470, 500, 530, 560, 610, \
		670, 770, 7000]$}\\\hline
		$450-600$&\makecell{$[100, 110, 120, 130, 140, 150, 160, 190, 210, 230, 250, 270, 290, 320, 340, 370, $\\$
		390, 420, 460, 490, 520, 550, 580, 620, 680, 780, 7000]$}\\\hline
		$600-800$&\makecell{$[100, 110, 120, 130, 150, 170, 200, 220, 240, 260, 280, 310, 340, 380, 410, 440, $\\$
		470, 510, 550, 620, 730, 7000]$}\\\hline
		$800-1100$&\makecell{$[100, 110, 120, 140, 160, 180, 200, 220, 250, 270, 300, 330, 360, 410, 460, 540, $\\$
		660, 7000]$}\\\hline
		$1100-1500$&\makecell{$[100, 130, 160, 190, 230, 270, 320, 400, 520, 7000]$}\\\hline
		$1500-2000$&\makecell{$[100, 210, 330, 7000]$}\\\hline
		$2000-2600$&\makecell{$[100, 7000]$}\\\hline
	\end{tabular}
    \caption{The fine binning where the relative statistical uncertainty of each bin must be smaller than $10\%$. The first column shows the ranges of the $m_{ll}$ bins, and the second column shows the boundaries of the $p_T$ bins. }
    \label{tab:bins_10pct}
\end{table}

\bibliographystyle{h-physrev}
\bibliography{DYg}

\end{document}